\documentclass[12pt,epsf,axodraw]{article}

\newcommand{\bea}{\begin{eqnarray}}
\newcommand{\eea}{\end{eqnarray}}
\newcommand{\be}{\begin{equation}}
\newcommand{\ee}{\begin{equation}}

\begin{document}

\thispagestyle{empty}

\begin{center}
{\LARGE \bf Truly Minimal Left-Right Model of\\ Quark and Lepton Masses\\}
\vspace{1.0in}
{\bf Biswajoy Brahmachari$^{1,2}$, Ernest Ma$^3$, and Utpal Sarkar$^4$
\\}

\vspace{0.5in}
{$^1$ \sl Theoretical Physics Division, Saha Institute of Nuclear Physics, \\
AF/1 Bidhannagar, Kolkata - 700 064, India} \\
\vspace{0.1in}
{$^2$ \sl Physics Department, Vidyasagar Evening College,\\
39, Sankar Ghosh Lane, Kolkata 700 006, India\\}
\vspace{0.1in}
{$^3$ \sl Physics Department, University of California, Riverside, California
92521, USA\\}
\vspace{0.1in}
{$^4$ \sl Theory Group, Physical Research Laboratory, Ahmedabad 380 009, India\\}
\vspace{0.1in}
\vspace{.5in}
\end{center}
\begin{abstract}
We propose a left-right model of quarks and leptons based on the gauge group 
$SU(3)_C \times SU(2)_L \times SU(2)_R \times U(1)_{B-L}$, where the scalar 
sector consists of only two doublets: (1,2,1,1) and (1,1,2,1).  As a result, 
any fermion mass, whether it be Majorana or Dirac, must come from 
dimension-five operators.  This allows us to have a common view of quark 
and lepton masses, including the smallness of Majorana neutrino masses as 
the consequence of a double seesaw mechanism.
\end{abstract}

In the standard model of electroweak interactions, neutrinos are massless.  
On the other hand, recent experimental data on atmospheric \cite{atm} and 
solar \cite{sol} neutrinos indicate strongly that they are massive and mix 
with one another \cite{others}. To allow neutrinos to be massive 
theoretically, the starting point is the observation of Weinberg 
\cite{wein79} over 20 years ago that a unique dimension-five operator 
exists in the standard model, i.e. 
\begin{equation}
{\cal L}_\Lambda = {f_{ij} \over 2 \Lambda} (\nu_i \phi^0 - e_i \phi^+) (\nu_j 
\phi^0 - e_j \phi^+) + H.c. \label{eq1}
\end{equation}
which generates a Majorana neutrino mass matrix given by
\begin{equation}
({\cal M}_\nu)_{ij} = {f_{ij} v^2 \over \Lambda}, \label{eq2}
\end{equation}
where $v$ is the vacuum expectation value of $\phi^0$.  This also shows 
that whatever the underlying mechanism for the Majorana neutrino mass, it 
has to be ``seesaw'' in character, i.e. $v^2$ divided by a large mass 
\cite{ma98}.

If the particle content of the standard model is extended to include 
left-right symmetry \cite{lr}, then the gauge group becomes ${\cal G}_{LR}
\equiv SU(3)_C \times SU(2)_L \times SU(2)_R \times U(1)_{B-L}$, whose 
diagonal generators satisfy the charge relationship
\begin{equation}
Q = T_{3L} + T_{3R} + \frac{(B-L)}{2} = T_{3L} + \frac{Y}{2}. \label{eq3}
\end{equation}
Quarks and leptons transform as
\begin{eqnarray}
q_L = (u,d)_L &\sim& (3,2,1,1/3), \label{eq04} \\ 
q_R = (u,d)_R &\sim& (3,1,2,1/3), \label{eq4} \\ 
l_L = (\nu,e)_L &\sim& (1,2,1,-1), \label{eq05} \\ 
l_R = (N,e)_R &\sim& (1,1,2,-1), \label{eq5}
\end{eqnarray}
where a new fermion, i.e.~$N_R$, has been added in order that the left-right 
symmetry be maintained.

In all previous left-right models, a scalar bidoublet transforming as 
$(1,2,2,0)$ is then included for the obvious reason that we want masses 
for the quarks and leptons.  Suppose however that we are only interested in 
the spontaneous breaking of $SU(2)_L \times SU(2)_R \times U(1)_{B-L}$ to 
$U(1)_{em}$ with $v_R >> v_L$, then the simplest way is to introduce two 
Higgs doublets transforming as
\begin{eqnarray}
\Phi_L = (\phi^+_L,\phi^0_L) &\sim& (1,2,1,1), \label{eq06} \\ 
\Phi_R = (\phi^+_R,\phi^0_R) &\sim& (1,1,2,1). \label{eq6}
\end{eqnarray}
Suppose we now do not admit any other scalar multiplet.  This is analogous 
to the situation in the standard model, where $SU(2)_L \times U(1)_Y$ is 
spontaneously broken down to $U(1)_{em}$ by a Higgs doublet and we do not 
admit any other scalar multiplet.  In that case, we find that quark and 
charged-lepton masses are automatically generated by the existing Higgs 
doublet, but neutrinos obtain Majorana masses only through the dimension-five 
operator of Eq.~(\ref{eq1}).  
In our case, in the absence of the bidoublet, all 
fermion masses, be they Majorana or Dirac, must now have their origin in 
dimension-five operators, as shown below.

Using Eqs.~(\ref{eq04}) to (\ref{eq6}), it is clear that
\begin{equation}
(l_L \Phi_L) = \nu_L \phi_L^0 - e_L \phi_L^+ \label{eq7}
\end{equation}
and
\begin{equation}
(l_R \Phi_R) = N_R \phi_R^0 - e_R \phi_R^+  \label{eq8}
\end{equation}
are invariants under ${\cal G}_{LR}$.  Hence we have the dimension-five 
operators given by
\begin{equation}
{\cal L}_M = {f^L_{ij} \over 2 \Lambda_M} (l_{iL} \Phi_L)(l_{jL} \Phi_L) 
+ {f^R_{ij} \over 2 \Lambda_M} (l_{iR} \Phi_R)(l_{jR} \Phi_R) + H.c.,
\label{eq9}
\end{equation}
which will generate Majorana neutrino masses proportional to $v_L^2/\Lambda_M$ 
for $\nu_L$ and $v_R^2/\Lambda_M$ for $N_R$.  (The different possible origins 
of this operator are explained fully in Ref.[5].) In addition, we have
\begin{equation}
{\cal L}_D = {f^D_{ij} \over \Lambda_D} (\bar l_{iL} \Phi_L^*)
(l_{jR} \Phi_R) + H.c.
\label{eq10}
\end{equation}
and the corresponding dimension-five operators which will generate Dirac 
masses for all the quarks and charged leptons.

From Eq.~(\ref{eq10}), it is clear that
\begin{equation}
(m_D)_{ij} = {f^D_{ij} v_L v_R \over \Lambda_D}, \label{eq11}
\end{equation}
hence $\nu_L$ gets a double seesaw \cite{dss} mass of order
\begin{equation}
{m_D^2 \over m_N} \sim {v_L^2 v_R^2 \over \Lambda_D^2} {\Lambda_M \over v_R^2} 
= {v_L^2 \Lambda_M \over \Lambda_D^2}, \label{eq12}
\end{equation}
which is much larger than $v_L^2/\Lambda_M$ if $\Lambda_D << \Lambda_M$.  
Take for example $\Lambda_M$ to be the Planck scale of $10^{19}$ GeV and 
$\Lambda_D$ to be the grand-unification scale of $10^{16}$ GeV, then the 
neutrino mass scale is 1 eV (for $v_L$ of order 100 GeV).  The difference 
between $\Lambda_M$ and $\Lambda_D$ may be due to the fact that if we assign 
a global fermion number $F$ to $l_L$ and $l_R$, then ${\cal L}_M$ has $F = 
\pm 2$ but ${\cal L}_D$ has $F=0$.

Since the Dirac masses of quarks and charged leptons are also given by 
Eq.~(\ref{eq11}), $v_R$ cannot be much below $\Lambda_D$.  This means 
that $SU(2)_R 
\times U(1)_{B-L}$ is broken at a very high scale to $U(1)_Y$, and our model 
at low energy is just the standard model.  We do however have the extra 
singlet neutrinos $N_R$ with masses of order $v_R^2/\Lambda_M$, i.e. below 
$10^{13}$ GeV, which are useful for leptogenesis, as is well-known \cite{lg}.

For $m_t = 174.3 \pm 5.1$ GeV, we need $v_R/\Lambda_D$ to be of order unity 
in Eq.~(\ref{eq11}).  One may wonder in that case whether we can still write 
Eq.~(\ref{eq10}) as an effective operator.  The answer is yes, as can be 
seen with the following specific example \cite{dwr}.  Consider the singlets
\begin{equation}
U_L, U_R \sim (3,1,1,4/3), \label{eq13}
\end{equation}
with invariant mass $M_U$ of order $\Lambda_D$, then the $2 \times 2$ 
mass matrix linking $(\bar t_L, \bar U_L)$ to $(t_R,U_R)$ is given by
\begin{equation}
{\cal M}_{tU} = \pmatrix{0 & f_t^L v_L \cr f_t^R v_R & M_U}.
\label{eq14}
\end{equation}
For $v_L << v_R, M_U$, we then have
\begin{equation}
m_t = {f_t^L f_t^R v_L v_R \over M_U} \left[ 1 + {(f_t^R v_R)^2 \over M_U^2} 
\right]^{-{1 \over 2}},
\label{eq15}
\end{equation}
which is in the form of Eq.~(14) even if $v_R/M_U \sim 1$.

Since we already have dimension-five operators, we should also consider 
dimension-six operators.  In that case, we can invoke the Bardeen-Hill-Lindner 
(BHL) dynamical mechanism \cite{bhl} with a cutoff scale equal to $\Lambda_D$. 
We may assume that 
the effective dynamical BHL Higgs doublet [call it $\Phi_1 = (\phi_1^+,
\phi_1^0)$] couples only to the top quark, whereas our fundamental $\Phi_L$ 
[call it $\Phi_2$] couples to all quarks and leptons.  We thus have a 
specific two-Higgs-doublet model \cite{daskao} with experimentally verifiable 
phenomenology, as described below.

Since the BHL model predicts $m_t = 226$ GeV for $\Lambda_D = 10^{16}$ GeV, 
the effective Yukawa coupling of $\bar t_L t_R$ to $\bar \phi_1^0$ is
\begin{equation}
f_t^{(1)} = (226~{\rm GeV}) (2\sqrt 2 G_F)^{1 \over 2} = 226/174 = 1.3,
\label{eq16}
\end{equation}
and for $\tan \beta = v_2/v_1$, we have
\begin{equation}
m_t = (1.3 \cos \beta + f_t^{(2)} \sin \beta)(174~{\rm GeV}).
\label{eq17}
\end{equation}
This shows that, with a second Higgs doublet, the correct value of $m_t$ may 
be obtained.  Furthermore, $f_t^{(2)}$ may be assumed to be small, say of 
order $10^{-2}$.  This allows $v_R/\Lambda_D \sim 10^{-2}$ in Eq.~(14) and 
thus $v_R \sim 10^{14}$ GeV, so that $m_N \sim v_R^2/\Lambda_M$ is of order 
$10^9$ GeV, which may be more effective for leptogenesis, even with the 
reheating of the Universe after inflation.  At the same time, using 
Eq.~(\ref{eq17}), this fixes
\begin{equation}
\tan \beta \simeq 0.83
\label{eq18}
\end{equation}
for the phenomenology of the two-doublet Higgs sector.

Since the $d,s,b$ quarks receive masses only from $v_2$, there is no 
tree-level flavor-changing neutral currents in this sector.  This explains the 
suppression of $K_L-K_S$ mixing and $B - \bar B$ mixing.  On the other hand, 
both $v_1$ and $v_2$ contribute to the $u,c,t$ quarks, so our model does 
predict tree-level flavor-changing neutral curents in this sector.  Suppose 
the Yukawa interaction $f_t^{(1)} \bar \phi_1^0 \bar t_L t_R$ is replaced 
by $f_t^{(1)} (v_1/v_2) \bar \phi_2^0 \bar t_L t_R$, then the resulting 
mass matrix would be exactly proportional to the Yukawa matrix.  This means 
that there would not be any flavor-nondiagonal interactions.  Hence the 
term which contains all the flavor-changing interactions is given by 
\cite{ma01}
\begin{equation}
f_t^{(1)} \left( \bar \phi_1^0 - {v_1 \over v_2} \bar \phi_2^0 \right) \bar 
t'_L t'_R + H.c.,
\label{eq19}
\end{equation}
where $t'_{L,R}$ are the original entries in the $u,c,t$ mass matrix before 
diagonalization to obtain the mass eigenstates.  We thus expect contributions 
to, say $D - \bar D$ mixing, beyond that of the standard model.  Let
\begin{equation}
t'_{L,R} \simeq t_{L,R} + \epsilon_{tc}^{L,R} c_{L,R} + \epsilon_{tu}^{L,R} 
u_{L,R},
\label{eq20}
\end{equation}
where the $\epsilon$ parameters are at most of order $f_t^{(2)}/f_t^{(1)} \sim 
10^{-2}$, then \cite{ma01}
\begin{equation}
{\Delta m_{D^0} \over m_{D^0}} \simeq {B_D f_D^2 [f_t^{(1)}]^2 \over 3 
m^2_{eff} \sin^2 \beta} |\epsilon^L_{tc} \epsilon^L_{tu} \epsilon^R_{tc} 
\epsilon^R_{tu}|,
\label{eq21}
\end{equation}
where $m^2_{eff}$ is the effective normalized contribution from $\phi_1^0 - 
(v_1/v_2) \phi_2^0$.  Using $f_D = 150$ MeV, $B_D = 0.8$, and the present 
experimental upper bound \cite{pdg} of $2.5 \times 10^{-14}$ on this fraction, 
we then obtain
\begin{equation}
{|\epsilon^L_{tc} \epsilon^L_{tu} \epsilon^R_{tc} \epsilon^R_{tu}| \over 
10^{-8}} \left( {100~{\rm GeV} \over m_{eff}} \right)^2 < 1.
\label{eq22}
\end{equation}
This shows that $D - \bar D$ mixing may be observable in this model, in 
contrast to the negligible expectation of the standard model.

Rare top decays such as $t \to c$ (or $u$) + neutral Higgs boson are now 
possible if kinematically allowed.  Their branching fractions are of order 
$|\epsilon|^2 \sim 10^{-4}$.  Once a neutral Higgs boson is produced at a 
future collider, its decay will also be a test of this model.  Its dominant 
decay is still $b \bar b$, but its subdominant decays will not just be 
$c \bar c$ and $\tau^- \tau^+$, but also $c \bar u$ and $u \bar c$.  There 
should be observable $D^\pm \pi^\mp$ final states, for example.

Since we want $v_R >> v_L$ in our model, it is advantageous to extend it 
to include supersymmetry to solve the hierarchy problem.  The quark and 
lepton multiplets of Eqs.~(4) to (7) are now superfields, and the scalar 
multiplets of Eqs.~(8) and (9) will have partners
\begin{equation}
\Phi_L^c \sim (1,2,1,-1), ~~~ \Phi_R^c \sim (1,1,2,-1),
\end{equation}
and all four are superfields as well.  All fermion masses must again be 
given by dimension-five operators, coming from the $(l_{iL} \Phi_L)(l_{jL} 
\Phi_L)$ terms in the superpotential, etc.  However, we need to impose an 
exactly conserved discrete symmetry: odd for all quark and lepton superfields, 
but even for the Higgs superfields to forbid the bilinear $(l_{iL} \Phi_L)$ 
terms, etc.  [This is equivalent to the usual $R$ parity of the minimal 
supersymmetric standard model (MSSM) even though we do not have trilinear 
terms at all in the superpotential of this model.]  With $SU(2)_R$ breaking 
at the $10^{16}$ GeV scale, the particle content of this model at low energy 
is identical to that of the 
MSSM with the requisite two light Higgs doublets.  This solves the hierarchy 
problem and is a well-known case of gauge coupling unification.

In conclusion, we have proposed a truly minimal $SU(3)_C \times SU(2)_L \times 
SU(2)_R \times U(1)_{B-L}$ gauge model, with the simplest possible Higgs 
sector.  All fermion masses, be they Majorana or Dirac, have a common origin, 
i.e. dimension-five operators.  Whereas Dirac fermions have masses at the 
electroweak scale, the observed neutrinos have naturally small Majorana 
masses from a double seesaw mechanism.  The existence of singlet right-handed 
neutrinos with masses in the range $10^9$ to $10^{13}$ GeV are required, and 
their decays establish a lepton asymmetry which is converted at the 
electroweak phase transition to the present observed baryon asymmetry of 
the Universe.

In the standard model, Yukawa couplings of quarks and leptons to Higgs 
doublets are renormalizable. This means that the fermion mass matrices are 
arbitrary at any scale.  In our case, the effective dimension-five operators 
have their origin at some high-energy scale.  The new physics there fixes the 
coefficients of those operators (perhaps according to some symmetry valid 
above that scale) in analogy to the seesaw mechanism which fixes the Majorana 
neutrino mass operator at that scale.  Below it, we have the standard model, 
so the evolution of the Yukawa couplings determine the charged fermion mass 
matrices at the electroweak scale.  This is a natural framework for a possible 
theoretical understanding of fermion masses and mixing in the future.

Since our proposed model is identical to the standard model below $10^{16}$ 
GeV (except for the $N_R$'s), or to the MSSM in the supersymmetric version, 
the usual predictions of the latter also apply, including the expected 
occurrence of proton decay and neutron-antineutron oscillations from 
higher-dimensional operators due to new physics at or above $10^{16}$ GeV.

In the presence of dimension-six operators, we may invoke the 
Bardeen-Hill-Lindner mechanism to generate a dynamical Higgs doublet which 
renders the $top$ quark massive.  Since we also have a fundamental Higgs 
doublet, this allows us to have a realistic $m_t$ (which is not possible 
in the minimal BHL model) and an effective two-doublet Higgs sector at 
the electroweak scale with distinctive and experimentally verifiable 
flavor-changing phenomena.\\[5pt]

The work of EM was supported in part by the U.~S.~Department of Energy
under Grant No.~DE-FG03-94ER40837. BB would like to thank the organisers
of the flavour@cern series of talks, where some aspects of the 
double seesaw mechanism were discussed.

\end{document}